\documentclass[%
superscriptaddress,
onecolumn,
floatfix,
]{revtex4-2}

\nonstopmode

\usepackage{dcolumn}%
\usepackage{bm}%
\usepackage{hyperref}%
\usepackage{float,afterpage}
\usepackage{amsmath,amssymb}
\usepackage{siunitx}  
\usepackage{lineno}
\usepackage{tabularx}
\usepackage{graphicx}
\usepackage{geometry}
\begin{document}

\title{Optical Magnetic Field Enhancement using Ultrafast Azimuthally Polarized Laser Beams and Tailored Metallic Nanoantennas}




\author{Rodrigo Mart\'in-Hern\'andez}
\thanks{These authors contributed equally to this work}
\affiliation{Grupo de Investigaci\'on en Aplicaciones del L\'aser y Fot\'onica, Departamento de F\'isica Aplicada, Universidad de Salamanca, E-37008, Salamanca, Spain.}

\author{Lorenz Gr\"unewald}
\thanks{These authors contributed equally to this work}
\affiliation{Institute of Theoretical Chemistry, Faculty of Chemistry, University of Vienna, 1090 Vienna, Austria}
\affiliation{Vienna Doctoral School in Chemistry (DoSChem), Faculty of Chemistry, University of Vienna, 1090 Vienna, Austria}

\author{Luis Sánchez-Tejerina}
\affiliation{Grupo de Investigaci\'on en Aplicaciones del L\'aser y Fot\'onica, Departamento de F\'isica Aplicada, Universidad de Salamanca, E-37008, Salamanca, Spain.}
\affiliation{Departamento de Electricidad y Electrónica, Universidad de Valladolid, 47011 Valladolid, Spain}

\author{Luis Plaja}
\affiliation{Grupo de Investigaci\'on en Aplicaciones del L\'aser y Fot\'onica, Departamento de F\'isica Aplicada, Universidad de Salamanca, E-37008, Salamanca, Spain.}
\affiliation{Unidad de Excelencia en Luz y Materia Estructuradas (LUMES), Universidad de Salamanca, Salamanca, Spain}

\author{Enrique Conejero Jarque}
\affiliation{Grupo de Investigaci\'on en Aplicaciones del L\'aser y Fot\'onica, Departamento de F\'isica Aplicada, Universidad de Salamanca, E-37008, Salamanca, Spain.}
\affiliation{Unidad de Excelencia en Luz y Materia Estructuradas (LUMES), Universidad de Salamanca, Salamanca, Spain}

\author{Carlos Hern\'andez-Garc\'ia}
\email{carloshergar@usal.es}
\affiliation{Grupo de Investigaci\'on en Aplicaciones del L\'aser y Fot\'onica, Departamento de F\'isica Aplicada, Universidad de Salamanca, E-37008, Salamanca, Spain.}
\affiliation{Unidad de Excelencia en Luz y Materia Estructuradas (LUMES), Universidad de Salamanca, Salamanca, Spain}

\author{Sebastian Mai}
\email{sebastian.mai@univie.ac.at}
\affiliation{Institute of Theoretical Chemistry, Faculty of Chemistry, University of Vienna, 1090 Vienna, Austria}





\begin{abstract} 
Structured light provides unique opportunities to spatially tailor the electromagnetic field of laser beams. 
This includes the possibility of a sub-wavelength spatial separation of their electric and magnetic fields, which would allow isolating interactions of matter with pure magnetic (or electric) fields.
This could be particularly interesting in molecular spectroscopy, as excitations due to electric and---usually very weak---magnetic transition dipole moments can be disentangled.
In this work, we show that the use of tailored metallic nanoantennas drastically enhances the strength of the longitudinal magnetic field carried by an ultrafast azimuthally polarized beam (by a factor of $\sim65$), which is spatially separated from the electric field by the beam's symmetry.
Such enhancement is due to favorable phase-matching of the magnetic field induced by the electronic current loops created in the antennas.
Our particle-in-cell simulation results demonstrate that the interaction of moderately intense ($\sim10^{11}$\,W/cm$^2$) and ultrafast azimuthally polarized laser beams with conical, parabolic, Gaussian, or logarithmic metallic nanoantennas provide spatially isolated magnetic field pulses of several tens of Tesla.
\end{abstract}
\maketitle
\section{Introduction} 

Current optical techniques allow us to tailor laser light in all spatiotemporal dimensions, opening the possibility to structure the electromagnetic field in unprecedented ways \cite{2019_forbes_Structured, 2022_He_structured}. 
One of the most intriguing developments, as highlighted in \cite{2023_bliokh_Roadmap}, lies in the ability to separate the electric field (E-Field) and the magnetic field (B-field) of a laser beam. 
This presents an opportunity to disentangle interactions between purely electric and purely magnetic field-matter interactions. 
There are several approaches to isolate (i.e., to spatially separate) the E-field and the B-field components of an optical laser beam. 
For example, close to mirror surfaces, interference effects lead to different positions of the nodes of the E-field and B-field, which, for example, affects the emission rate of electronic and magnetic dipole transitions of molecules close to the surface
\cite{2011_karaveli_Spectral, 2016_rabouw_EuropiumDoped}.
Similar effects are observed near dielectric or metal photonic nano-structures
\cite{2018_sanz-paz_Enhancing, 2015_hussain_Enhancing}, in cavities (referred to as Purcell effect) \cite{1946_purcell_Resonance}, or in metasurfaces \cite{2015_veysi_metasurfaces}.
A potentially simpler approach consists of using two counter-propagating beams, where the E-field and B-field nodes can be separated spatially; such a scheme has been proposed, e.g., with high-power laser beams with applications in attoscience \cite{2023_Martin-Hernandez}.

\begin{figure}[tb]
    \centering
    \includegraphics[width=.9\linewidth]{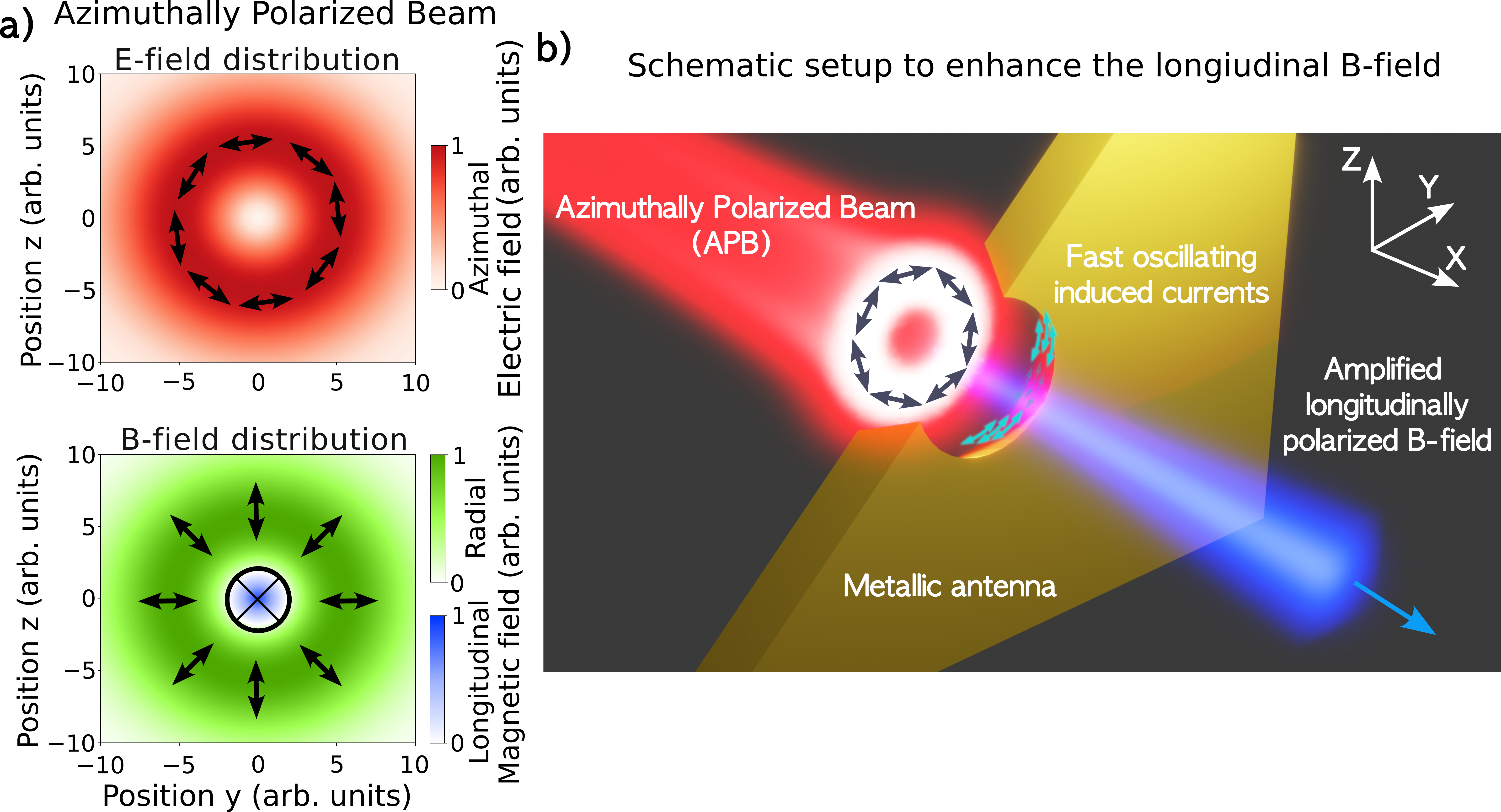}
    \caption{a) Azimuthally polarized beam (APB) electric field (E-field, upper panel) and magnetic field (B-field, lower panel) spatial distributions. 
    The arrows in each panel indicate the polarization direction.
    The B-field radial polarization distribution is depicted in green, while the blue corresponds with the longitudinal component (out of plane). 
    b) The APB (red beam) interacts with a metallic nanoantenna, inducing strong and fast oscillating currents in the inner surface of the antenna (light blue, double headed arrows). 
    As a result of the coherent superposition of the magnetic fields created by the ultrafast induced currents, the longitudinally polarized magnetic field (blue beam) of the APB is amplified by more than an order magnitude.}
    \label{fig:Scheme}
\end{figure}

Alternatively, spatially isolated E-fields and B-fields can be achieved with structured light techniques~\cite{2019_forbes_Structured}, which allow precisely sculpting the spatial distribution of amplitude, phase, and polarization state of a laser beam.
One prominent example of such structured light beams are vector beams, laser beams with spatially dependent polarization states \cite{2021_chekhova_Polarization}. 
The most common vector beams are radially polarized and azimuthally polarized beams (APB), which show a cylindrically symmetric linearly polarized pattern \cite{2009_zhan_Cylindrical}. 
They are of fundamental interest in applications such as optical communication technology, nano-lithography, and quantum key distribution
\cite{2017_rubinsztein-dunlop_Roadmap, 2023_bliokh_Roadmap, 2023_Shen_roadmap}. 
In an APB, the B-field is polarized radially, whereas the E-field is azimuthally (i.e., tangentially) polarized.
This polarization state acts similar to an oscillating closed current loop, thus inducing a longitudinally polarized B-field along the beam's propagation axis upon focusing of the APB.
Given that the E-field vanishes on the propagation axis due to symmetry constraints (i.e., it shows a singularity), in an APB the longitudinal B-field is locally isolated from the E-field (see Fig.~\ref{fig:Scheme}a). 
Such isolated B-fields have been proposed as a tool for several applications, e.g., magnetic spectroscopy \cite{Veysi2016, 2002_zurita-sanchez_Multipolar, 2021_wozniak_Single} or force microscopy \cite{Zeng2018, 2010_rajapaksa_Image} employing sub-Tesla B-fields, or ultrafast nonlinear magnetization \cite{2023_Sanchez-Tejerina} employing multi-Tesla B-fields.

The possibility to isolate intense B-fields provides interesting opportunities in optical spectroscopy. 
A considerable fraction of our knowledge related to electronic states of atoms and molecules has been obtained by spectroscopic investigations at optical wavelengths, e.g., by UV/Vis absorption spectroscopy.
At optical wavelengths, the interaction between the E-field component of the absorbed light and the molecular transition electric dipole (ED) moment dominates \cite{2009_burresi_Probing, 2012_taminiau_Quantifying, 2012_bernadotte_Originindependent, 2015_kasperczyk_Excitation}, whereas the interaction of the B-field component of light with the molecular transition magnetic dipole (MD) moment is orders of magnitude weaker \cite{2005_foot_Atomic}. 
Although weak, MD transitions are potentially of strong interest in the elucidation of ``dark'' states which cannot easily be accessed or observed via strong ED transitions (e.g., parity-conserving transitions in metal atoms or $n\pi^*$ states)
\cite{1925_laporte_Rules, 2016_rabouw_EuropiumDoped, 2015_parson_Modern, Clouthier_1983, Klessinger1995, 1949_platt_Classification, 2018_valiev_Relations}.
In such systems, the direct observation of MD transitions would enable a more detailed investigation of the electronic/vibronic structure of the target molecules, as well as investigations of photoinduced processes initiated in such states.
In order to observe MD transitions, which are typically concealed by the much stronger ED transitions, special strategies are required.
Circular dichroism (CD) or magnetic circular dichroism (MCD) are both sensitive to MD transitions, but are only applicable to certain kinds of molecular systems (chiral molecules or ones that exhibit certain near-degeneracies, respectively) \cite{2003_mason_Magnetic}.
Hence, it would be desirable to have a general approach that does not only work in certain systems (e.g., those with very large MD moments or forbidden ED transitions).
As discussed previously in the literature, such a general approach would require that the E-field is strongly suppressed at the target molecule's position \cite{2015_kasperczyk_Excitation}.
Additionally, due to the intrinsic weakness of MD transitions, it would be advantageous to employ a B-field that is as strong as possible.
The combination of such spatially isolated and intense oscillating B-fields could then form the basis for a general ``MD-only optical spectroscopy''.

APBs are excellent candidates for this purpose, as recently demonstrated by Kasperczyk et al. \cite{2015_kasperczyk_Excitation}, who employed a focused APB to record a luminescence excitation spectrum of Eu$^{3+}$-doped Y$_2$O$_3$ nanoparticles. 
In particular, at the center of the beam, emission was only detected when the wavelength was tuned to the relevant MD transition of Eu$^{3+}$, whereas no emission was observed when exciting the ED transition.
This shows that suppression of the E-field can be used to observe MD-exclusive transitions. 
However, as MD transitions are intrinsically weak, a further enhancement of the B-field strength, beyond what is achievable by a focused APB, would be desirable.
A simultaneous isolation and enhancement of the B-field would not only enable MD-only spectroscopy, but also provide opportunities for other applications, e.g., in laser-driven particle acceleration \cite{2015_rassou_Influence,2017_korneev_Magnetization,2011_vieira_Magnetic}, photoinduced force microscopy \cite{2010_rajapaksa_Image, 2015_huang_Imaging}, or other nanoscale force microscopy techniques \cite{2000_hecht_Scanning, 1995_zenhausern_Scanning}.

The generation of APBs in the visible and near-infrared can be achieved with the use of spatial light modulators \cite{Beversluis_2006}, uniaxial or biaxial crystals \cite{2013_Shvedov, 2015_Turpin}, optical fibers \cite{Hirayama_2006}, or with nano-structured half-waveplates (known as s-waveplates) \cite{2011_beresna_Polarization}, which modify the polarization state distribution of the incident light beam accordingly. 
A very attractive possibility to generate \textit{intense} APBs ---and thus intense and isolated B-fields---is to move to the ultrafast regime. 
Recent experiments have shown the possibility of generating vector beams in the near-infrared/femtosecond scale using s-waveplates \cite{2020_alonso_Complete}, or even in the extreme ultraviolet/attosecond regime through high harmonic generation \cite{2017_Hernandez-garcia_VB}. 
But most notably, the combination of ultrafast vector beams and nanoantennas, where the longitudinal B field is enhanced through the creation of oscillating current loops, has pioneered the generation of isolated Tesla-scale magnetic fields \cite{blanco_2019, Sederberg_2020}, ideal candidates for ultrafast MD-only spectroscopy and other applications.

In this work, we explore the use of specificallly tailored nanoantennas---rotationally symmetric, metallic nano-structures (see Fig.~\ref{fig:Scheme}b)---to enhance the spatially isolated longitudinal B-field of an ultrafast APB. 
We perform particle-in-cell (PIC) simulations, using the OSIRIS code \cite{2002_fonseca_OSIRIS, 2008_fonseca_Onetoone, 2013_fonseca_Exploiting}, to simulate and analyze the spatiotemporal evolution of an ultrafast APB interacting with metallic nanoantennas of cylindrical \cite{blanco_2019} and various other shapes.
To understand the underlying physics, we develop an analytical model based on the retarded potential formalism \cite{2017_griffiths_Introduction,jackson1999classical}, which allows us to explain the enhancement mechanism in terms of favourable phase-matching between ultrafast electronic currents. 
Expanding upon the previous simulations \cite{blanco_2019}, we investigate the capabilities of sloped and curved antennas---conical, parabolic, Gaussian, and logarithmic---with varying geometric parameters for B-field enhancement and simultaneous E-field suppression of incident APBs. 
We also show how the enhanced B-field spatial distribution depends on the antenna shape.
Our work provides a rigorous study of how to enhance and tailor the spatial distribution of the B-field carried by an ultrafast APB through the use of custom metal nanoantennas in the optical, femtosecond regime.

\section{Methods}

In our simulations, we predict the temporal and three-dimensional spatial evolution of ultrafast APBs with the goal of enhancing and isolating its longitudinal B-field. 
To do so, we explore its interaction with metallic antennas with different shapes and geometry parameters.

\subsection{Characterization of an ultrafast azimuthally polarized laser beam}
\label{sec:analytical_apb_expression}

Throughout our simulations, we employ an APB as our incident laser beam.
We formulate our incident beam at the entrance plane of the simulation box in the paraxial approximation, propagating in a vacuum along the $x$ direction.
In cylindrical coordinates $\vec{r}=\left(\rho,\phi,x\right)^\textrm{T}$, with the focus placed at the origin ($x=0$), the electromagnetic field of such an APB can be written as:
\begin{linenomath}\begin{align}
    \vec{E}(\vec{r}, t) 
    &= 
    E_0 f(t)e^{\text{i}\frac{2\pi}{\lambda} \left( x-ct\right)}\frac{\sqrt{2}}{w_0}\rho 
    \text{e}^{-\frac{\rho^2}{w_0^2}}\vec{u}_\phi ,
    \label{eq:apb_efield} 
    \\ 
    \vec{B}(\vec{r}, t) 
    &= 
    -\frac{E_0}{c} f(t)e^{\text{i}\frac{2\pi}{\lambda} \left( x-ct\right)}\frac{\sqrt{2}}{w_0}
    \text{e}^{-\frac{\rho^2}{w_0^2}} 
    \left(
    \rho \vec{u}_{\rho}
    + \frac{\lambda}{\pi} 
    \left(1 - \frac{\rho^2}{w_0^2}\right)
    \text{e}^{\text{i}\frac{\pi}{2}}\vec{u}_{x}\right),
    \label{eq:apb_bfield}  
\end{align}\end{linenomath}
where $w_0$ is the beam waist, $\lambda$ is the central wavelength, and $c$ is the vacuum speed of light. 
$\vec{u}_{\rho}$, $\vec{u}_{\phi}$, $\vec{u}_{x}$ are unit polarization vectors, and $f(t)=\sin^2(\pi\frac{t}{\tau_\mathrm{end}})$ is the temporal envelope function for $0 \leq t \leq \tau_\mathrm{end}$.
To alleviate the high computational cost of simulations, we employed a short laser pulse with a pulse duration $\tau_\mathrm{end} = 10\:\text{fs}$.

The spatial distributions of the E-field and B-field intensities at the focal plane are represented in Fig.~\ref{fig:Scheme}a.  
In contrast to a fundamental Gaussian beam, the E-field peak amplitude $E_\text{max}=E_0/\sqrt{\mathrm{e}}$ 
is not reached at the minimum waist $w_0$ but instead at $\rho_\textrm{max}=w_0/\sqrt{2}$.
Also note that the radially polarized ($\vec{u}_\rho$) component of the B-field spatially coincides with the azimuthally polarized ($\vec{u}_\phi$) E-field.
Directly on-axis ($\rho=0$), all transversal ($\vec{u}_\rho$ and $\vec{u}_\phi$) E-field and B-field components vanish, and only the longitudinally polarized B-field ($\vec{u}_x$) remains non-zero, reproducing the above-mentioned E-field--B-field separation.
The latter component increases with decreasing waist parameter $w_0$, i.e., upon focusing \cite{2016_veysi_Focused}.

Within our PIC simulations (see details below), we set the minimum waist to $w_0 = 2.5\,\mathrm{\mu m}$ and the E-field peak amplitude to $E_\text{max}=\,1.2\,\mathrm{GV/m}$ (equivalent to $E_0=\,1.98\,\mathrm{GV/m} $ and a peak intensity of $1.9\times10^{11}\,\mathrm{W/cm^2}$ at $\rho_\mathrm{max}$).
With the above parameters, the longitudinal B-field peak amplitude at the focus of the beam is $B_\text{max}(\rho=0,x=0) = 0.6$\,T. 
The central wavelength is set at $\lambda=527.5$\,nm, motivated by the excitation wavelength of the prominent ${}^7F_0\rightarrow{}^5D_1$ MD-exclusive absorption band in the Eu$^{3+}$ ion,\cite{1997_bihari_Spectra, 2015_binnemans_Interpretation} which was used previously to demonstrate exclusive MD excitation via an APB \cite{2015_kasperczyk_Excitation}.


\subsection{Simulation framework}
We propagate the described APB numerically using the Maxwell solver implemented in the PIC package OSIRIS \cite{2002_fonseca_OSIRIS, 2008_fonseca_Onetoone, 2013_fonseca_Exploiting}.
Here, the E-fields and B-fields are represented on a 3D mesh.
In our simulations, we describe the metallic antenna material as spatially restricted neutral plasma, which within PIC is described by charged macroparticles that move according to Newton's equation and are self-consistently coupled to the E-field and B-field propagation \cite{1983_dawson_Particle, 2004_birdsall_Plasma, 2018_derouillat_Smilei}.

OSIRIS internally works with a suitably normalized system of dimensionless units, which can be scaled as desired a posteriori.
In order to relate our simulations to the parameters mentioned above (minimum waist, peak amplitude, central wavelength), we chose a reference frequency with a value of $\omega_\mathrm{R}=2\pi c/527.5\,\mathrm{nm}$.

The neutral plasma representing the antenna material is made up of two particle species, motivated by the presence of electrons and gold ions in metallic gold.
For both species, we consider a particle species density $n_e = n_n = 5.9\times10^{22}\,\mathrm{cm^{-3}}$ (assuming monovalent gold ions).
The mass-to-charge ratio for the gold ions was set to ${m_{Au}}/{q_{Au}} = 361630\: m_e/e$ in terms of the electron mass.
Note that such a neutral plasma cannot accurately represent material parameters such as the frequency-dependent electric permittivity or magnetic permeability. 
However, given that our total pulse duration (10 fs) is below the room temperature carrier relaxation time of the metallic antenna (27.3\,fs in Au, $\sim$ 36\,fs in other metals as Ag or Cu  \cite{Gall2016}), the material properties can be approximated to a free electron model, described by this neutral plasma. 
For larger pulse durations, relaxation and damping effects should be taken into account for a realistic description.


\subsection{Antenna shapes and geometry parameters}

\begin{figure}[b] 
    \centering
    \includegraphics[width=.8\linewidth]{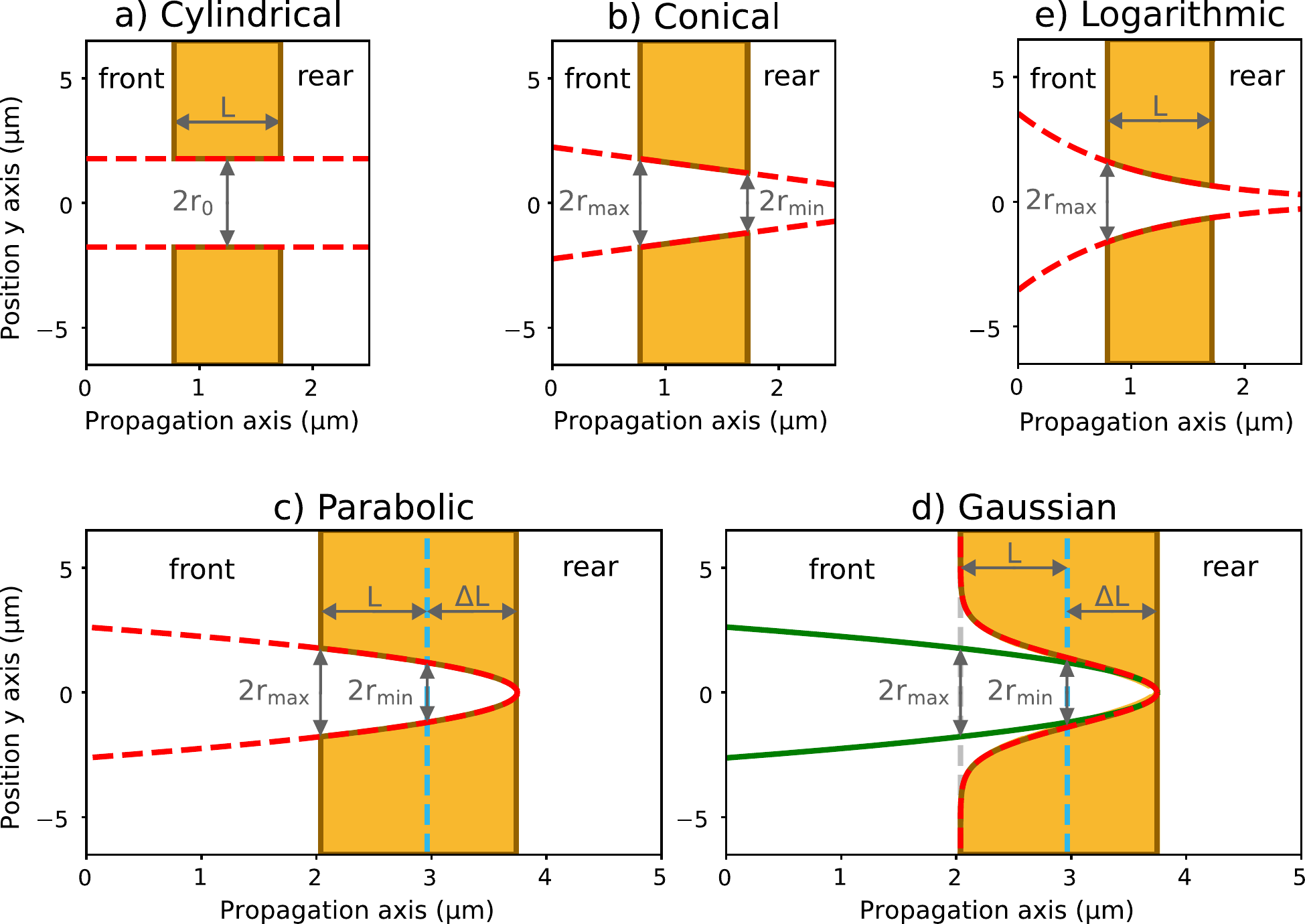} 
    \caption{
    Vertical section of the considered antenna shapes (all antennas are cylindrically symmetric around the propagation axis). 
    The orifice dimensions are governed by analytical functions (see Tab.~\ref{tab:geometries}, indicated by the red dashed lines. 
    The Gaussian antenna case is parametrized to fit to the parabola shape (green).
    The shapes are labeled a--e in the order they are discussed below.
    }
    \label{fig:aperture_shapes}
\end{figure}

In this work we explore the B-field enhancement and isolation in five different antennas: cylindrical, conical, parabolic, Gaussian, and logarithmic. 
The investigated shapes all exhibit cylindrical symmetry around the optical axis of the laser beam.
Note that deviations from this symmetry (i.e., using an elliptical aperture) were shown to diminish the B-field enhancement \cite{blanco_2019}.
The profiles for the proposed antennas are shown in Fig.~\ref{fig:aperture_shapes}a--e, and the corresponding parameters are detailed in Tab.~\ref{tab:geometries}, in Appendix~\ref{apx:AntennaShapes}.

In our simulations for conical, cylindrical, and logarithmic antennas we use a box size of $l_x=2.5\,\mu\textrm{m}$ (longitudinal) and $l_y=l_z=13\,\mu\textrm{m}$ (transversal).
With the employed omnidirectional spatial resolution of 9.0\,nm, this yields a grid of $279\times1445\times1445$ cells.
For the parabolic and Gaussian shapes, the longitudinal box size was doubled to $l_x=5.0\,\mu$m  (with $558\times1445\times1445$ cells). 
This was necessary to retain the B-field maximum inside the simulation box for the wide set of geometry parameters that we consider.
All antennas are centered in the longitudinal direction at the beam focus, which is placed at $x_0=l_x/2$. 
The simulation time was set to 15\,fs for the conical, cylindrical and logarithmic antennas, and to 30\,fs for the parabolic and Gaussian ones, with a temporal resolution of 5.6\,as, which is well within the limits given by the Courant-Friedrichs-Lewy (CFL) condition \cite{1966_yee_Numerical}.
For all walls of the simulation box, we employ perfectly matched layers with 90\,nm thickness as absorbing boundary conditions.\cite{2000_vay_New}

The OSIRIS input files and evaluation scripts are made publicly available at Zenodo.\cite{grunewald_2023_10082669}

\section{Results and Discussion}

In this section, we start with an analysis of the B-field enhancement upon the optimization of the antenna geometries with a linear orifice slope along the optical axis, i.e., cylindrical and conical antennas. 
Thereafter, we introduce an analytical model that allows us to understand the physics beyond the longitudinal B-field enhancement in a cylindrical antenna. 
Then, we extend our study to antenna geometries that present a non-linear profile along the optical axis, i.e., parabolic, Gaussian, and logarithmic antennas.
Finally, we analyze the contrast between B-fields and E-fields intensities for all proposed antenna geometries to quantify the spatial isolation of the intense longitudinal B-fields.


\subsection{Magnetic field enhancement for linearly sloped antennas}

The simplest antenna shapes are the cylindrical and the conical antennas. Here we present PIC numerical simulations that show the B-field enhancement in these cases.


\subsubsection{Cylindrical antenna}

In a previous study by Blanco et al.\cite{blanco_2019}, cylindrical metallic antennas were proposed for enhancing the B-field of an APB.
The study investigated in detail the influence of the antenna radius $r_0$ on the B-field enhancement.
It was found that the highest B-field is obtained when $r_0=\rho_\text{max}$, i.e., when the antenna radius coincides with the radial position where the E-field of an APB reaches its maximum amplitude (see Eq.~\eqref{eq:apb_efield}). 

\begin{figure}
    \centering
    \includegraphics[width=\textwidth]{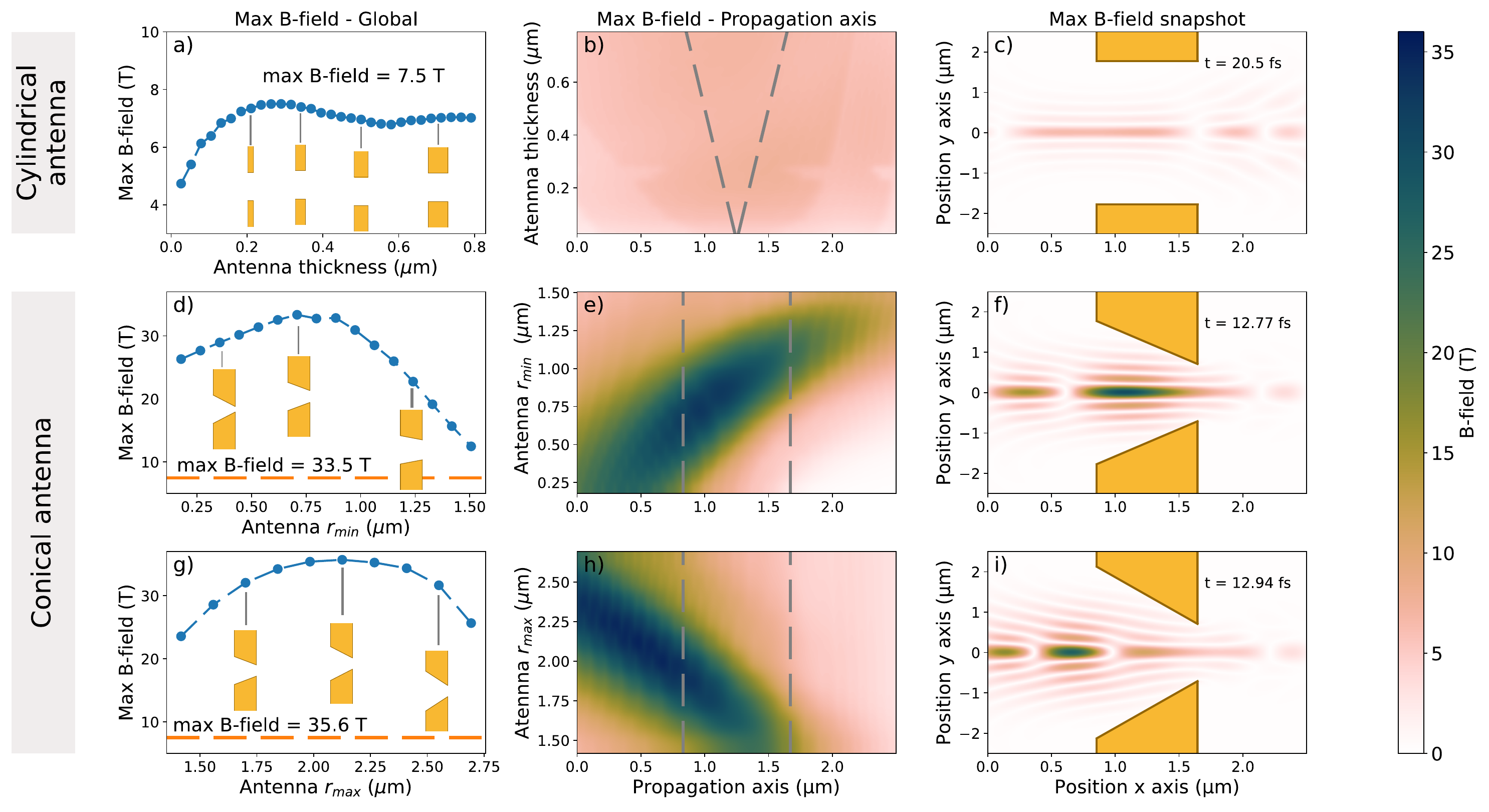}
    \caption{Maximum achievable longitudinal B-field strength for different antenna thicknesses (a-c) of the cylindrical antenna, for the $r_\textrm{min}$ scan (d-f), and for the $r_\textrm{max}$ scan of the conical antenna. 
    In the first column (a,d,g), the maximum longitudinal B-field on the propagation axis is shown along the scanned parameter, together with the B-field of the optimized cylindrical antenna (orange dashed line). 
    The second column (b,e,h) reveals the distribution of the B-field strength along the propagation axis, the longitudinal extent of the aperture (grey dashed lines). 
    In the third column (c,f,i), a temporal snapshot of the longitudinally polarized B-field, when the maximum amplitude is reached, is presented for the different scans.} 
    \label{fig:ConeScanBfield} 
\end{figure}

Here, we expand upon these investigations exploring the role of the antenna thickness $L$ (considering  $r_0=\rho_\textrm{max}$).
As presented in Fig.~\ref{fig:ConeScanBfield}a, the maximum B-field strength grows from $4.5\,\mathrm{T}$ for the thinnest antenna, with $L=0.03\,\mathrm{\mu m}$, to $7.5\,\mathrm{T}$ that is reached at $L=0.3\,\mathrm{\mu m}$. 
For larger $L$, the maximum B-field strength roughly oscillates between $6.5$ and $7.5\,\mathrm{T}$. 
Our simulations then show that once a minimum thickness of $L=0.3\,\mathrm{\mu m}$ is achieved, thicker antennas do not provide higher longitudinal B-fields but also do not diminish the B-field strength; thus, the thickness can be adapted to practical needs, e.g., thermal robustness. 

In Fig.~\ref{fig:ConeScanBfield}b, we analyze the B-field strength along the propagation axis ($x$) for different values of the antenna thickness $L$.
It can be seen that similar maximum B-fields are obtained over a wide range of positions before, within, and behind the antenna (depicted by the grey dashed lines).
The needle-shaped B-field distribution for a particular case ($L=0.3\,\mathrm{\mu m}$) is represented in Fig.~\ref{fig:ConeScanBfield}c.
Unlike the large spatial extent of the B-field needle along the propagation axis, the B-field strength drops quickly (over tens of nanometers) when moving away from the optical axis in the transverse direction.
Consequently, experiments relying on the isolated and enhanced B-field in a cylindrical antenna  would require precise placement of the target particles/molecules.

\subsubsection{Conical antenna}
\label{sec:conical_antenna}
The saturation observed in Fig.~\ref{fig:ConeScanBfield}a motivated us to modify the geometry to a conically-shaped antenna, in order to maximize the direct interaction between the APB and the metal.
The conical antenna geometry is a generalization of the cylindrical antenna that exhibits a linear slope.
We assume an open conical antenna facing the incident beam, i.e., with front radius $r_\textrm{max}$ and rear radius $r_\textrm{min}$ such that $r_\textrm{max}>r_\textrm{min}>0$. 
The antenna thickness $L$ is fixed at $L=0.8\,\mathrm{\mu m}$, the largest value we investigated for the cylindrical antenna.

In order to find values of $r_\textrm{max}$ and $r_\textrm{min}$ that optimize the B-field strength, in Fig.~\ref{fig:ConeScanBfield}d we first scan $r_\textrm{min}$ while keeping $r_\textrm{max}$ fixed at the optimal radius for cylindrical aperture ($r_0=1.77\,\mathrm{\mu m}$).
Note that the far right of the figure ($r_\textrm{min}=r_0=1.77\,\mathrm{\mu m}$) corresponds to the cylindrical case.
If $r_\textrm{min}$ is then reduced, we can observe a significant increase in the maximum B-field strength, reaching a peak value of $33.5\,\mathrm{T}$ for $r_\textrm{min} \sim 0.75\,\mathrm{\mu m}$. 
This translates into a gain factor of $\sim50$ compared to the freely propagating focused APB without any antenna, and a gain factor of $\sim5$ compared to the optimized cylindrical antenna (orange line).
In this case, the B-field is longitudinally focused in a relatively narrow region, as exhibited in the spatial distributions shown in Figs.~\ref{fig:ConeScanBfield}e and f, as long as $r_\textrm{min}$ is clearly smaller than $r_\textrm{max}$.
The maximum B-field is found in the rear of the antenna position for large $r_\textrm{min}$ (i.e., for nearly cylindrical antennas) and moves in front of the antenna for decreasing $r_\textrm{min}$ (i.e., for more strongly sloped walls), as depicted in Fig.~\ref{fig:ConeScanBfield}e.
The highest achieved B-field strengths are obtained when the B-field peak is located inside the antenna ($0.85\,\mathrm{\mu m}<x<1.65\,\mathrm{\mu m}$).
This shows that by varying $r_\textrm{min}$---or analogously, the slope of the antenna walls---we can tune both the gain factor and the B-field peak position (i.e., we can switch from a reflective to a transmissive antenna), which is potentially of high interest for the envisioned applications.

In order to fully optimize the B-field enhancement in the interaction of an APB with conical antenna, we show in Fig.~\ref{fig:ConeScanBfield}g a scan over the front radius (${r_\textrm{max}}$) for a fixed rear radius of ${r_\textrm{min}} = 0.75 \,\mathrm{\mu m}$.
The optimal value is obtained for $r_\textrm{max}=2.13\,\mathrm{\mu m}$, which is higher than $r_0$, and further increases the B-field strength to $35.6\,\mathrm{T}$ (which conveys a gain factor of $\sim55$ compared to the free beam simulation).
Fig.~\ref{fig:ConeScanBfield}h shows the position along the propagation axis of the maximum B-field depending on ${r_\textrm{max}}$. 
We note that larger ${r_\textrm{max}}$ (more steeply sloped walls) move the B-field peak position towards the front part of the antenna.
The spatial snapshot presented in Fig.~\ref{fig:ConeScanBfield}i, shows that the optimal set of parameters ($L = 0.8 \,\mathrm{\mu m}$, ${r_\textrm{min}} = 0.75 \,\mathrm{\mu m}$, ${r_\textrm{max}} = 2.13 \,\mathrm{\mu m}$) lead to a localized B-field peak, located slightly before the antenna.


\subsection{Analytical model}
\label{sec:analytical_model}

To gain a better understanding of the physics underlying the interaction of an APB and a general conical antenna, we develop an analytical model based on the retarded potential formalism\cite{2017_griffiths_Introduction, jackson1999classical} (see complete derivation in Appendix~\ref{apx:AnalyticalModel}). 
The total longitudinal B-field ($B_x$) generated by a thick antenna is modelled as the coherent superposition of the fields ($B_{\text{loop}}$) generated by several, infinitesimally thin current loops induced in the metallic antenna inner surface.
As the total B-field ($B_x$) arises from this superposition, the phase difference between each $B_{\text{loop}}$ plays a major role in the final B-field enhancement. This is in analogous to the phase matching effect in nonlinear optics \cite{kurtz1968}. 
Thus, in this model the total longitudinal B-field, $B_x$, can be expressed in cylindrical coordinates by the following integral,
\begin{linenomath}\begin{align}
\label{eq:long_bfield}
    \begin{split}
    B_x\left(x,t\right) &= \int \mathrm{d}x' |B_\text{loop}(x,x')|\text{e}^{\textrm{i}\left(\phi(x,x')-\frac{2\pi c}{\lambda} t\right)}
    \\
    &= 2\pi\int \mathrm{d}x'\frac{j_0(x')}{c}\frac{\rho(x')^2}{\rho(x')^2 + (x-x')^2 }\left[\left(\frac{2\pi }{\lambda}\right)^2 + \frac{1}{\rho(x')^2 + (x-x')^2}\right]^{1/2}
    \\
    &\hspace{10em}
    \times \text{e}^{ \text{i} \frac{2\pi}{\lambda}\left( x' + \sqrt{\rho(x')^2 + (x-x')^2} -ct\right)  - \text{i}\arctan\left(\frac{2\pi}{\lambda}\sqrt{\rho(x')^2+(x-x')^2}\right)},
    \end{split}
\end{align}\end{linenomath}
where $j_0(x')$ is the current density distribution, and $\rho(x')$ is the aperture radius at $x'$. 
The variable $x'$ is the longitudinal position of the infinitesimally thin current loops, over which integration is carried out, along the propagation axis.

The spatial phase term $\phi(x,x')$ in Eq.~\eqref{eq:long_bfield} can be divided in two terms, $\phi(x,x') = \phi_1(x,x') + \phi_2(x,x')$.
The first phase term $\phi_1(x,x') = \frac{2\pi}{\lambda}\left(x' + \sqrt{\rho(x')^2 + (x-x')^2}\right)$ represents a fast spatial oscillation experienced by the B-field when propagating from the source point in the current loop to the observation point on the optical axis. 
The second term, $\phi_2(x,x') = -\arctan\left(\frac{2\pi}{\lambda}\sqrt{\rho(x')^2+(x-x')^2}\right)$, is a Gouy phase-like term, being minimal at the current loop centre ($x=x'$), and increasing up to $\frac{\pi}{2}$ for $|x-x'|\gg 0$, i.e., far from the current loop origin. The main contribution to the phase-matching-like process is given by the first phase term $\phi_1$, as $\phi_2$ remains almost constant in our region of interest.

\begin{figure}[htb]
    \centering
    \includegraphics[width=0.8\textwidth]{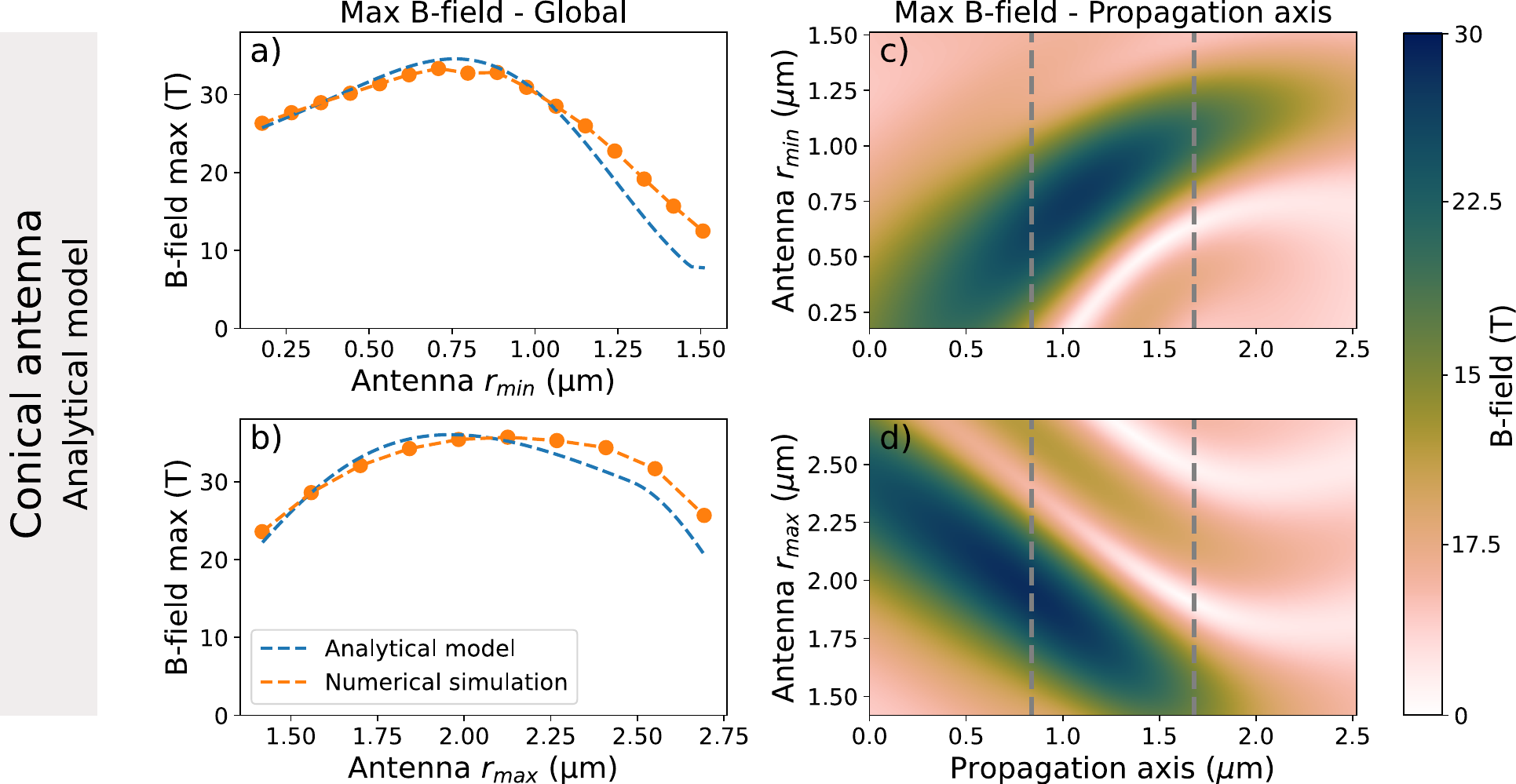}
    \caption{(a,b) Total B-field strength for cone orifice radius scans of the rear side ($r_\textrm{min}$ (a)) and the front side ($r_\textrm{max}$ (b)). The dotted orange line denotes for the results retrieved from numerical PIC simulations, whereas the blue dashed line illustrates the predicted B-field strength based on the derived analytical model (see Eq.~\ref{eq:long_bfield}). (c,d) Distribution of the analytically predicted B-field strength along the optical axis for cone orifice radius scans of the rear side ($r_\textrm{min}$ (c)) and the front side ($r_\textrm{max}$ (d)).}
    \label{fig:analytical_model_cone}
\end{figure}

The proposed analytical model is compared with the simulation results for the conical antenna in  Fig.~\ref{fig:analytical_model_cone}.  
Panels~a and b show the comparison of the maximum B-field strength of the previous two scans already presented in Figs.~\ref{fig:ConeScanBfield}d and g, whereas panels~c and d show the spatial extent of the maximum analytical B-field strength, analogous to Figs.~\ref{fig:ConeScanBfield}e and h.

The excellent agreement between the analytical model and the numerical simulations allows us to identify the physics underlying the B-field enhancement mechanism. According to Eq.~\eqref{eq:long_bfield}, the longitudinal B-field strength on axis is affected by the antenna shape in two ways: through the current density distribution $j_0(x')$, and through the accumulated phase $\phi$.
On the one hand, the current density distribution is given by $j_0(x')=q^2n\frac{E_0(x')}{\sqrt{2}m_e w}\left(1-\cos\left(\arctan\left(\frac{\partial \rho(x')}{\partial x'}\right)\right)\right)$, as detailed in Appendix~\ref{apx:AnalyticalModel}.
The dependence on the slope of the antenna walls ($\frac{\partial \rho(x')}{\partial x'}$) is such that the current density is largest if the incident beam is perpendicular to the wall, which increases the B-field strength in steep sloped conical antennas.
On the other hand, the B-field strength is controlled by the constructive interference given by $\phi(x,x')$. Taking into account the prominent role of $\phi_1$, optimal constructive interference would require a parabolic antenna shape, which shows that, in general, a curved antenna  optimizes the maximum B-field strength.

Thus, thanks to our analytical model we can conclude that the mechanism of B-field enhancement in the conical antenna is mostly based on the constructive interference of the local B-fields created by the transverse current loops distributed along the propagation axis. Depending on the slope of the antenna, this effect may confine the longitudinal B-field towards the front or the rear side of the antenna (see Figs.~\ref{fig:analytical_model_cone}b and d).


\subsection{Magnetic field enhancement for curved antennas}
 
As motivated in the previous section, further B-field enhancement can be achieved by introducing curved antenna shapes that optimize the constructive interference of the B-field along the propagation axis.
An additional advantage of curved/smooth antenna profiles is that they are possibly easier to manufacture\cite{2014_kollmann_Plasmonics, 2013_schoen_Planar, 2021_li_Recent} than antennas with perfectly straight walls.
Thus, in this section, we investigate the interaction of an APB with parabolic, Gaussian, and logarithmic antennas, with the ultimate goal of controlling the enhancement of the longitudinal B-field, and its confinement along the propagation axis.


\subsubsection{Parabolic antenna}
\label{sec:ParabolicAntenna}

Parabolic reflectors or mirrors are broadly used in different technological applications such as acoustics, optics, or telecommunications to confine incoming electromagnetic waves at one spatial point, the focus \cite{2019_penketh_Optimal}.
While such antennas usually have macroscopic dimensions, there has been growing interest in using nano-sized parabolas as small-scale optical antennas \cite{2019_penketh_Optimal, 2013_schoen_Planar, 2011_stockman_Nanoplasmonics}.
Within the discussion of the analytical model for the conical antenna in Section \ref{sec:analytical_model}, we showed that (near-) optimal constructive interference could be reached with a parabolic shape at its focal point.

The shape of a (truncated) parabolic antenna can be described with three independent parameters (see Fig.~\ref{fig:aperture_shapes}c, Table \ref{tab:geometries}).
First, the thickness $L$ represents the distance between the front and rear sides of the antenna (analogous to the previous antennas).
We fix the front-side parabola radius, $r_\textrm{max}$, which is located at $x_0-L/2$, to be at $r_\textrm{max}=w_\textrm{0}\sqrt{2}=1.77\,\mathrm{\mu m}$, as in the previous scans.
The second independent parameter is the rear-side radius, $r_\textrm{min}$, that we locate at $x_0+L/2$.
Then, $L$ and $r_\textrm{min}$ fix the focal length of the antenna, $f=\frac{r_\text{max}^2-r_\text{min}^2}{4L}$, which shows that $r_\textrm{min}$ controls the curvature of the antenna.
In order to investigate the B-field enhancement if the rear side of the antenna is closed, we introduce a third parameter, the cap thickness $\Delta L$, which shifts the rear wall of the antenna to $x_0+L/2+\Delta L$, while keeping its focal length and focus position fixed.

\begin{figure}
    \centering
    \includegraphics[width=\textwidth]{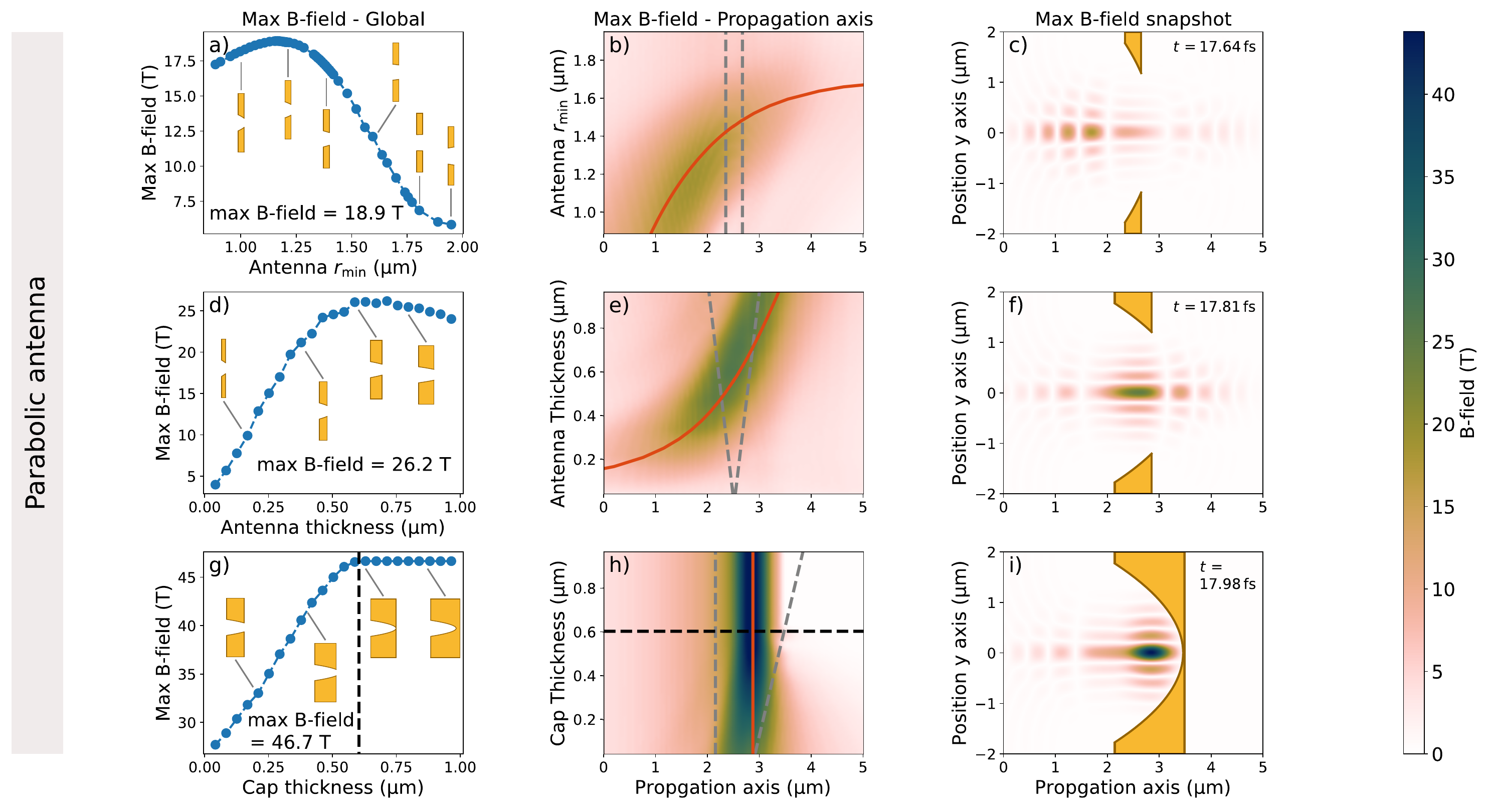}
    \caption{
    Absolute B-field strength for the parabolic antenna for scans of $r_\textrm{min}$ (a-c), antenna thickness $L$ (d-f) and cap thickness $\Delta L$ (g-i). 
    In the first column (a,d,g), the maximum B-field on the propagation axis is shown depending on the scanned parameter. 
    The second column (b,e,h) shows the distribution of the B-field strength along the propagation axis, the longitudinal extent of the aperture (grey, dashed lines), and the position of the focal point (red, solid line). 
    In the third column (c,f,i), a temporal snapshot of the highest B-field strength is presented for the different scans.
    The dashed black line in (g) and (h) indicates when the antenna is closed.
    }
    \label{fig:ParabolicScanBfield}
\end{figure}

The numerical simulation results that cover scans over the three parameters considered in a parabolic antenna are presented in Fig.~\ref{fig:ParabolicScanBfield}. 
First, panels ~a-c show the scan of $r_\textrm{min}$ over the range $0.89\,\mathrm{\mu m}\leq 1.94\,\mathrm{\mu m}$ (i.e., from $0.5r_\textrm{max}$ to $1.1r_\textrm{max}$).
A maximum on-axis B-field strength of 18.5\,T is reached for $r_\mathrm{min}\approx 1.17\,\mathrm{\mu m}$, which is about half the minimal waist.
The scan shows a rather broad peak, with 90\% of the maximum B-field strength reached between $1\,\mathrm{\mu m}$ and $1.4\,\mathrm{\mu m}$.
While for small $r_\textrm{min}$, the parabola is opening towards the front, for $r_\textrm{min}=r_\textrm{max}$ the antenna becomes cylindrical, and for $r_\textrm{min}>r_\textrm{max}$ opens towards the rear.
This increase of $r_\textrm{min}$ is accompanied by a substantial B-field strength weakening, where a rear-opened parabola enhances the B-field even less than a cylindrical antenna.
Moreover, the position of the maximum B-field shifts from the front side to the rear side along the scan, roughly following the focal point position, as shown in Fig.~\ref{fig:ParabolicScanBfield}b (red line).
Panel~c shows the B-field distribution for the optimal $r_\textrm{min}$ parameter, showing a sub-wavelength spot of maximum B-field at about $x=1.8\,\mathrm{\mu m}$, in front of the antenna.

Fig.~\ref{fig:ParabolicScanBfield}d-f presents the scan of the antenna thickness ($0.04\,\mathrm{\mu m}\leq L \leq 0.97\,\mathrm{\mu m}$), fixing the optimal values $r_\textrm{max}=1.77\,\mathrm{\mu m}$ and $r_\textrm{min}=1.17\,\mathrm{\mu m}$ from the previous scan.
For small $L$, the maximum B-field strength increases approximately linearly.
A maximum B-field of 25.4\,T is reached at $L=0.7\,\mathrm{\mu m}$, and a slow decrease of maximum B-field is observed beyond that point.
For the optimal parameters ($r_\textrm{min}$, $r_\textrm{max}$, $L$), the maximum B-field spot is located within the antenna, as shown in Fig.~\ref{fig:ParabolicScanBfield}e.
However, the maximum B-field spot can be shifted to the front or rear to some extent without compromising a strong B-field enhancement by varying $L$.
The respective needle-like B-field distribution for the optimal antenna thickness is illustrated in Fig.~\ref{fig:ParabolicScanBfield}f.

Finally, in Fig.~\ref{fig:ParabolicScanBfield}g, we scan the cap thickness, $\Delta L$, to investigate the B-field enhancement for closed parabolic antennas. 
The leftmost point is identical to the maximum in Fig.~\ref{fig:ParabolicScanBfield}d, and we scan over the range $0.0\,\mathrm{\mu m}\leq \Delta L\leq 1.0\,\mathrm{\mu m}$.
Considering that the APB does not exhibit strong E-fields close to the beam axis, one would not expect significant additional ring currents and thus no strong B-field enhancement when closing the antenna.
However, the scan in Fig.~\ref{fig:ParabolicScanBfield}g evidences a strong, linear increase in B-field from 25\,T to $\sim$47\,T.
Once the antenna is fully closed (at $\Delta L=0.6\,\mathrm{\mu m}$), no further increase in B-field strength is observed.
This increase in maximum B-field strength is accompanied by a strong confinement of the B-field, as seen in Fig.~\ref{fig:ParabolicScanBfield}h.
The same panel also shows that the B-field is zero behind the closed antenna, as expected.
The strong confinement is also visible in Fig.~\ref{fig:ParabolicScanBfield}i, demonstrating that the magnetic needle is compressed to a spot by the parabolic aperture.

Overall, we demonstrate that the closed parabolic antenna shape can yield B-field strengths well above those achieved with the cylindrical and conical antenna shapes, though the B-field is localized inside the antenna.
We revealed that---even in the near- or subwavelength regime where geometrical optics breaks down---the maximum B-field spot roughly coincides with the focal point of the parabolic reflector.
Both properties can be of fundamental interest if B-fields should be tightly confined at predefined positions in an experimental setup.

\subsubsection{Gaussian antenna}
\label{sec:GaussiancAntenna}

Although a parabolic antenna shape can yield a large B-field enhancement, the precise fabrication of its nano-sized structure is challenging \cite{2019_penketh_Optimal}.
Therefore, next we investigate Gaussian-shaped antennas, where the curved slope approximates a parabolic curve to second order near the beam axis.
Noticeably, nano-sized Gaussian-shaped holes in metal surfaces can be manufactured experimentally, e.g., via focused ion beam edging \cite{2015_vladov_Apparent, 2022_madison_Unmasking}. 

\begin{figure}
    \centering
    \includegraphics[width=\textwidth]{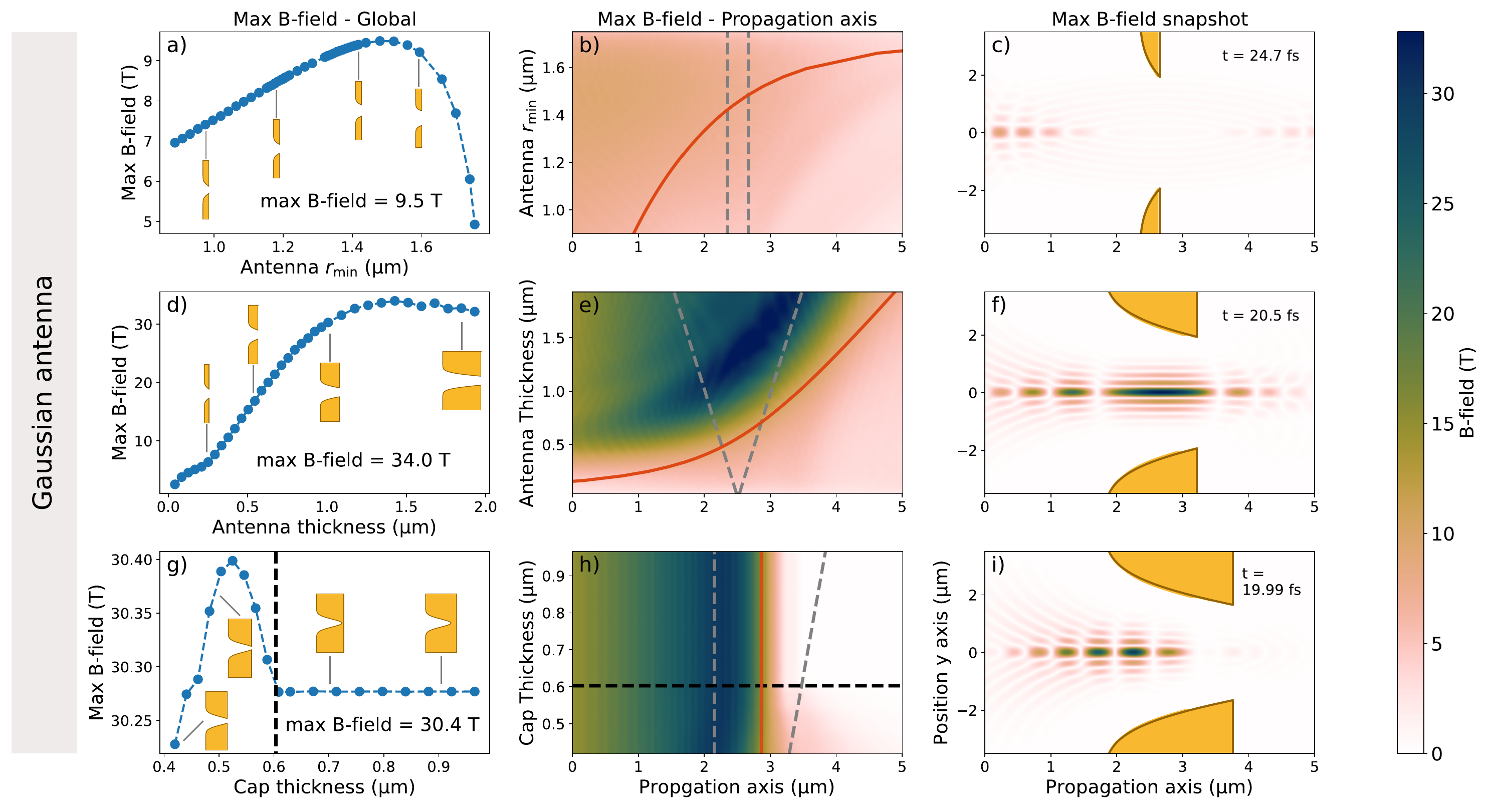}
    \caption{
    Absolute B-field strength for the Gaussian antenna for scans of $r_\textrm{min}$ (a-c), antenna thickness $L$ (d-f) and cap thickness $\Delta L$ (g-i). 
    In the first column (a,d,g), the maximum B-field on the propagation axis is shown depending on the scanned parameter. 
    The second column (b,e,h) shows the distribution of the B-field strength along the propagation axis, the longitudinal extent of the aperture (grey, dashed lines), and the position of the focal point (red, solid line). 
    In the third column (c,f,i), a temporal snapshot of the highest B-field strength is presented for the different scans.
    The dashed black line in (g) and (h) indicates when the antenna is closed.
    }
    \label{fig:GaussianScanBfield}
\end{figure}

Our choice for the parameter scans in the (truncated) Gaussian antenna follow closely the ones for the parabolic antenna (see Fig.~\ref{fig:aperture_shapes}c and d for comparison).
We performed three scans, for the rear radius $r_\textrm{min}$ at $x_0+L/2$, the antenna thickness $L$, and the cap thickness $\Delta L$.
The Gaussian baseline is located at $x_0-L/2$ and the Gaussian amplitude was set so that the Gaussian fits the parabola with the same set of parameters (compare Fig.~\ref{fig:aperture_shapes}d).

In Fig.~\ref{fig:GaussianScanBfield}a, we scan over the rear radius, $0.89\,\mathrm{\mu m}\leq r_\textrm{min}\leq 1.75\,\mathrm{\mu m}$.
For small $r_\textrm{min}$, the maximum B-field strength rises linearly, before reaching an optimal value of 9.5\,T at $r_\textrm{min}=1.5\,\mathrm{\mu m}$. 
For larger $r_\textrm{min}$, the maximum B-field strength falls of rapidly within few tens of nanometers, showing a different behaviour than observed for the parabolic antenna.
Panel~b shows that for the Gaussian antenna, the enhanced B-field is broadly distributed and does not follow the focal point of the parabola fitting the Gaussian (red line).
This can be rationalized as only the region close to the vertex resembles a concave parabolic shape, but further away from the beam axis changes to a convex shape that does not confine the B-field.
Consistent with the low maximum B-field strength and the large spread of the B-field, Fig.~\ref{fig:GaussianScanBfield}c shows no needle-like or spot-like confinement of the B-field.

The antenna length scan of the Gaussian antenna is shown in Fig.~\ref{fig:GaussianScanBfield}d.
The rear radius $r_\textrm{min}$ was fixed to the value that maximized the B-field for the parabolic antenna.
For thin Gaussian antennas ($L<0.3\,\mathrm{\mu m}$) the B-field strength grows slowly, while at thicker antenna lengths ($0.3,\mathrm{\mu m}\leq r_\textrm{min}\leq 1.2\,\mathrm{\mu m}$) we observe a faster increase in B-field strength.
The maximum B-field strength in this scan, 32\,T, is reached at about $L=1.45\,\mathrm{\mu m}$, which, surprisingly, exceeds the maximum B-field strength from the analogous scan in Fig.~\ref{fig:ParabolicScanBfield}d (25\,T at $L=0.7\,\mathrm{\mu m}$).
As depicted in Fig.~\ref{fig:GaussianScanBfield}e, the maximum B-field spot is located in front of the hypothetical focal point, independent of $L$.
This indicates that the Gaussian antenna with the chosen parameters does not behave like the fitted parabola.
The snapshot of the most intense B-field (Fig.~\ref{fig:GaussianScanBfield}f) within all simulations of the performed scan reveals that the maximum B-field is built up inside the aperture.

In the cap thickness ($\Delta L$) scan shown in Fig.~\ref{fig:GaussianScanBfield}g, the same parameters were fixed and scanned, respectively, as in the analogous scan of the parabolic antenna.
Remarkably, the maximum B-field is nearly constant over the scanned cap thickness, showing a value of about 30\,T for nearly closed and fully closed antennas.
This value is considerably smaller than the maximum B-field reached for the parabolic antenna---which reached up to 47\,T (Fig.~\ref{fig:ParabolicScanBfield}g)---but nonetheless shows a substantial enhancement of the B-field.
In Fig.~\ref{fig:GaussianScanBfield}h, we see that the maximum B-field spot is located on the front side of the antenna, and that no B-field reaches the vertex of the Gaussian profile.
Overall, the B-field enhancement in a Gaussian antenna is not as advantageous as in the parabolic antenna, where the constructive interference is more favorable, as evidenced in Fig.~\ref{fig:GaussianScanBfield}i.

\subsubsection{Logarithmic antenna}
\label{sec:LogarithmicAntenna}

Our final considered antenna geometry presents a logarithmic profile (see Fig.~\ref{fig:aperture_shapes}e).
In contrast to the Gaussian antenna profile, it exhibits a globally convexly curved profile. 
The front radius is fixed to $r_\textrm{max}=1.77\,\mathrm{\mu m}$ (as in all previous computations), which leaves two independent parameters (see Tab.~\ref{tab:geometries}), the decay parameter $\kappa$ and the antenna length $L$.

\begin{figure}
    \centering
    \includegraphics[width=\textwidth]{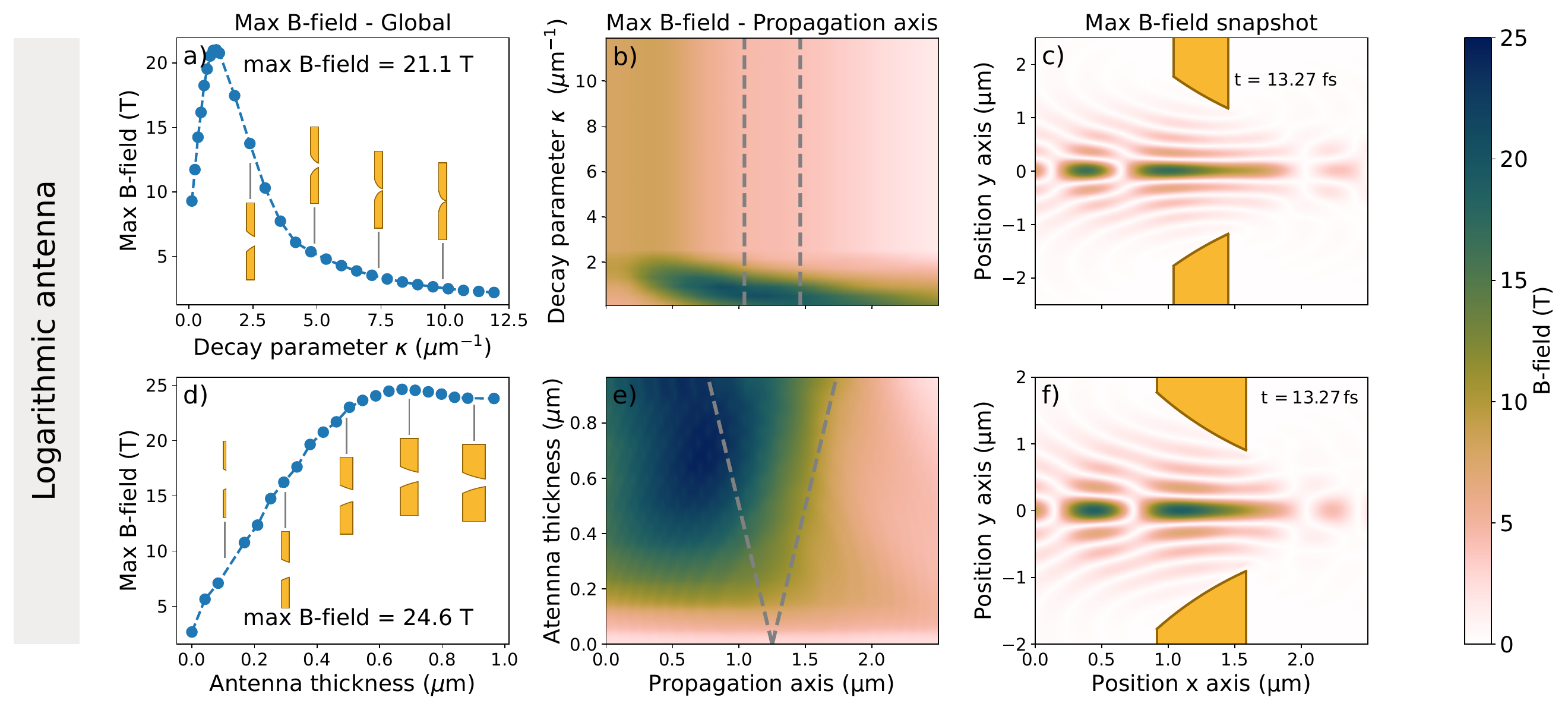}
    \caption{
    Absolute B-field strength for the logarithmic antenna for scans of $\kappa$ (a-c) and antenna thickness $L$ (d-f).
    In the first column (a,d) the maximum B-field on the propagation axis is shown depending on the scanned parameter. 
    The second column (b,e) shows the distribution of the B-field strength along the propagation axis and the longitudinal extent of the aperture (grey, dashed lines).
    In the third column (c,f), a temporal snapshot of the highest B-field strength is presented for the different scans.
    }
    \label{fig:LogarithmicScanBfield}
\end{figure}

In Fig.~\ref{fig:LogarithmicScanBfield}a, the antenna decay parameter is scanned in the interval $0.12,\mathrm{\mu m^{-1}}\leq \kappa\leq 12\,\mathrm{\mu m^{-1}}$.
The antenna thickness was fixed at $L=0.42\,\mathrm{\mu m}$.
We observe a maximum B-field peak around $\kappa = 1.15\,\mathrm{\mu m^{-1}}$ with a B-field strength of 21\,T.
For smaller decay parameters, the maximum achievable B-field quickly drops of, until the logarithmic antenna resembles the cylindrical one at $\kappa = 0.0\,\mathrm{\mu m^{-1}}$
For $\kappa$ values larger than the optimum, the rear antenna radius ($r_\text{min}$) decreases (exponentially) until the antenna is nearly closed, leading to a strong reduction in the maximum achievable B-field that is even lower than for free beam propagation.
As shown in Fig.~\ref{fig:LogarithmicScanBfield}b, at small $\kappa$ values the (open) antenna works similar to the other presented antennas with a maximum B-field spot whose position (rear to front) depends sensitively on $\kappa$.
At larger $\kappa$ values, the antenna is nearly closed, and does not enhance the B-field.
In the optimal case, for $\kappa = 1.15\,\mathrm{\mu m^{-1}}$, the maximum B-field spot is located inside the antenna, close to the front edge. 
The maximum B-field spot position is more readily visible in the snapshot in Fig.~\ref{fig:LogarithmicScanBfield}c.

In the antenna thickness scan shown in Fig.~\ref{fig:LogarithmicScanBfield}d, the parameters $\kappa = 1.15 \,\mathrm{\mu m^{-1}}$ and $r_\textrm{max}=1.77\,\mathrm{\mu m}$ were kept fixed, and the thickness was varied in the interval $L=0.04\,\mathrm{\mu m}$ to $L=0.97 \, \mathrm{\mu m}$.
The maximum B-field strength grows approximately linearly between $L=0.04\,\mathrm{\mu m}$ and about $L=0.5\,\mathrm{\mu m}$, reaching its peak at $L=0.68\,\mathrm{\mu m}$ with a B-field strength of about 25\,T.
For thicker antennas, the maximum B-field strength stays approximately constant.
As shown in Fig.~\ref{fig:LogarithmicScanBfield}~e, the maximum B-field position is always in front of the antenna, only weakly depending on $L$.
It can be observed that the B-field in the rear is always very weak, and thus we do not expect that a further increase in $L$ would change the B-field distribution.
The B-field snapshot shown in Fig.~\ref{fig:LogarithmicScanBfield}f is very similar to the one presented in Fig.~\ref{fig:LogarithmicScanBfield}c, except for a slightly higher maximum B-field.

Overall, the logarithmic antenna can be compared most closely to the conical antenna.
The results show that for similar parameters ($r_\text{max}$, effective $r_\text{min}$, $L$), the conical antenna provides a larger B-field enhancement.
The maximum B-field achieved with the logarithmic antenna is about 22\,T, whereas the conical antenna was able to deliver up to about 36\,T (which is about 50\% larger).
This shows that already a small (convex) curvature of the antenna wall can substantially decrease the effectiveness of the device.

\subsection{Magnetic-to-electric field contrast}

Up to now we have analyzed the B-field enhancement of an APB with a metallic nanoantenna. 
However, for applications not only the intensity of the B-field is important, but also the spatial separation between the E-field and B-field components.
For example, in an MD-only optical spectroscopy scheme, MD transitions can only be unambiguously distinguished from ED and electric quadrupole transitions if the E-field is effectively suppressed\cite{2015_kasperczyk_Excitation, 2019_cohen-tannoudji_Quantenmechanika}.
To examine the isolation of the B-field from the E-field, we compute the intensity contrast, which we define as the $c$-scaled ratio of the square-integrated B-field and E-field, $c^2\bar{B}^2/\bar{E}^2$.
The quantities $\bar{B}^2$ and $\bar{E}^2$ denote the temporally integrated B-field and E-field intensities, defined as
\begin{linenomath}\begin{equation}
    \bar{E}^2 = \frac{1}{T}\int_0^T |E(\vec{r},t)|^2 \mathrm{d}t,
    \qquad
    \bar{B}^2 = \frac{1}{T}\int_0^T |B(\vec{r},t)|^2 \mathrm{d}t,
\end{equation}\end{linenomath}
where $T$ is the total simulation time.

\begin{figure}[tb]
    \centering
    \includegraphics[width=8.5cm]{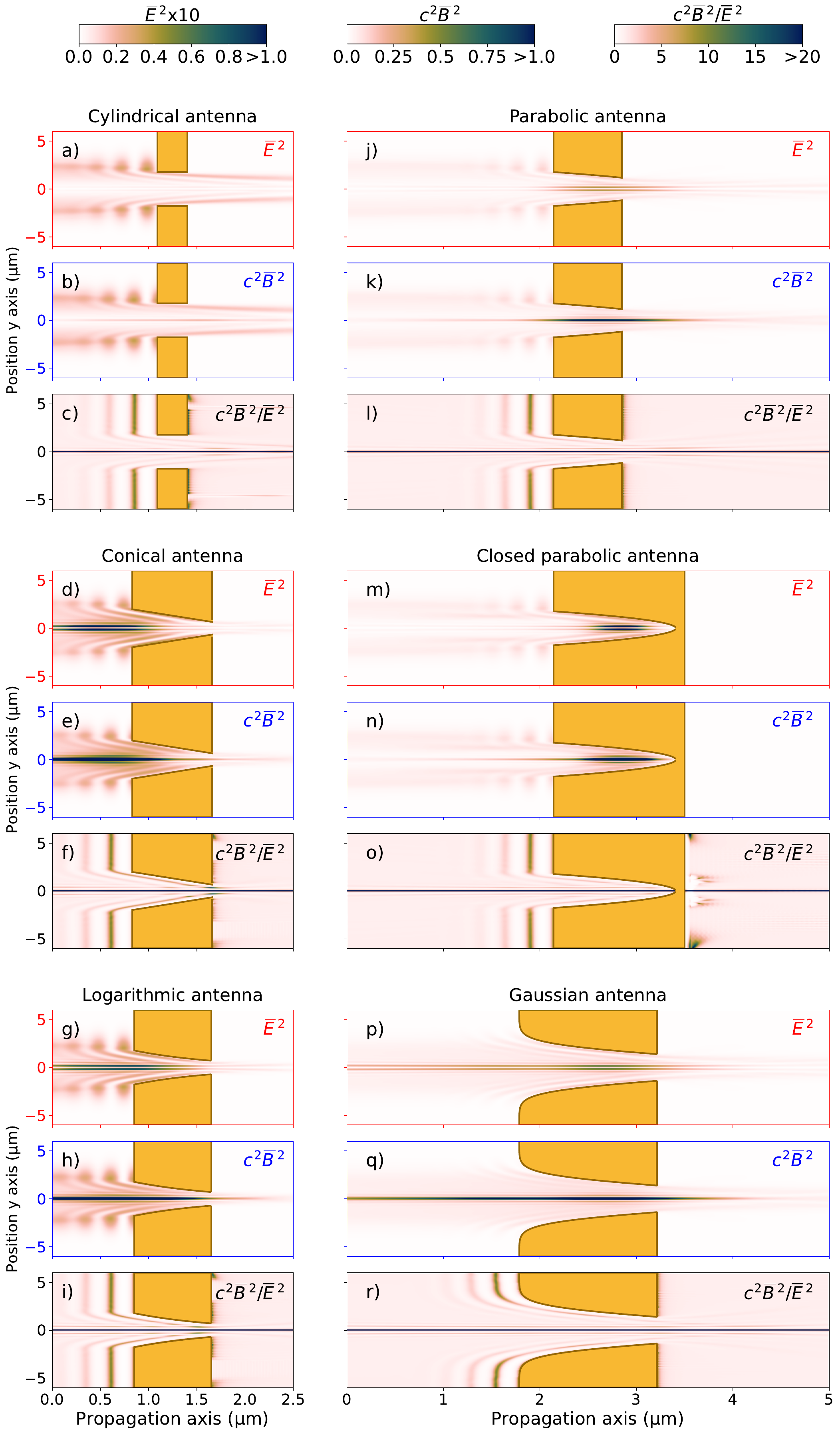}
    \caption{
    Temporally integrated squared-integrated E-field/B-field (red/blue) and intensity contrast $\frac{c^2\bar{B}^2}{\bar{E}^2}$ for the (a-c) cylindrical, (d-f) conical, (g-i) logarithmic, (j-l) parabolic, (m-o) closed parabolic, and (p-r) Gaussian antenna geometries. 
    For each antenna, the optimal parameters (see above) were used.
    }
    \label{fig:contrast_heatmap}
\end{figure}
In Fig.~\ref{fig:contrast_heatmap} we present the intensity contrast for the different aperture shapes considered in this work, using the optimal parameters mentioned above. 
First, we observe transverse regions of high contrast due to the interference between the incoming and reflected fields. 
Defining a desired contrast of 10 (for which the B-field intensity is one order of magnitude higher than the E-field intensity), these regions exhibit a thickness of about $x_c \approx 50\,\mathrm{nm}$, about one tenth of the wavelength.
Note that at the rear side of the antennas, the E-field and B-field are negligible, so the contrast is ill-defined.
Otherwise, the different analyzed geometries show quite similar intensity contrast distributions, with only minor differences near the antenna edges.

\begin{figure}[b]
    \centering
    \includegraphics[width=0.9\textwidth]{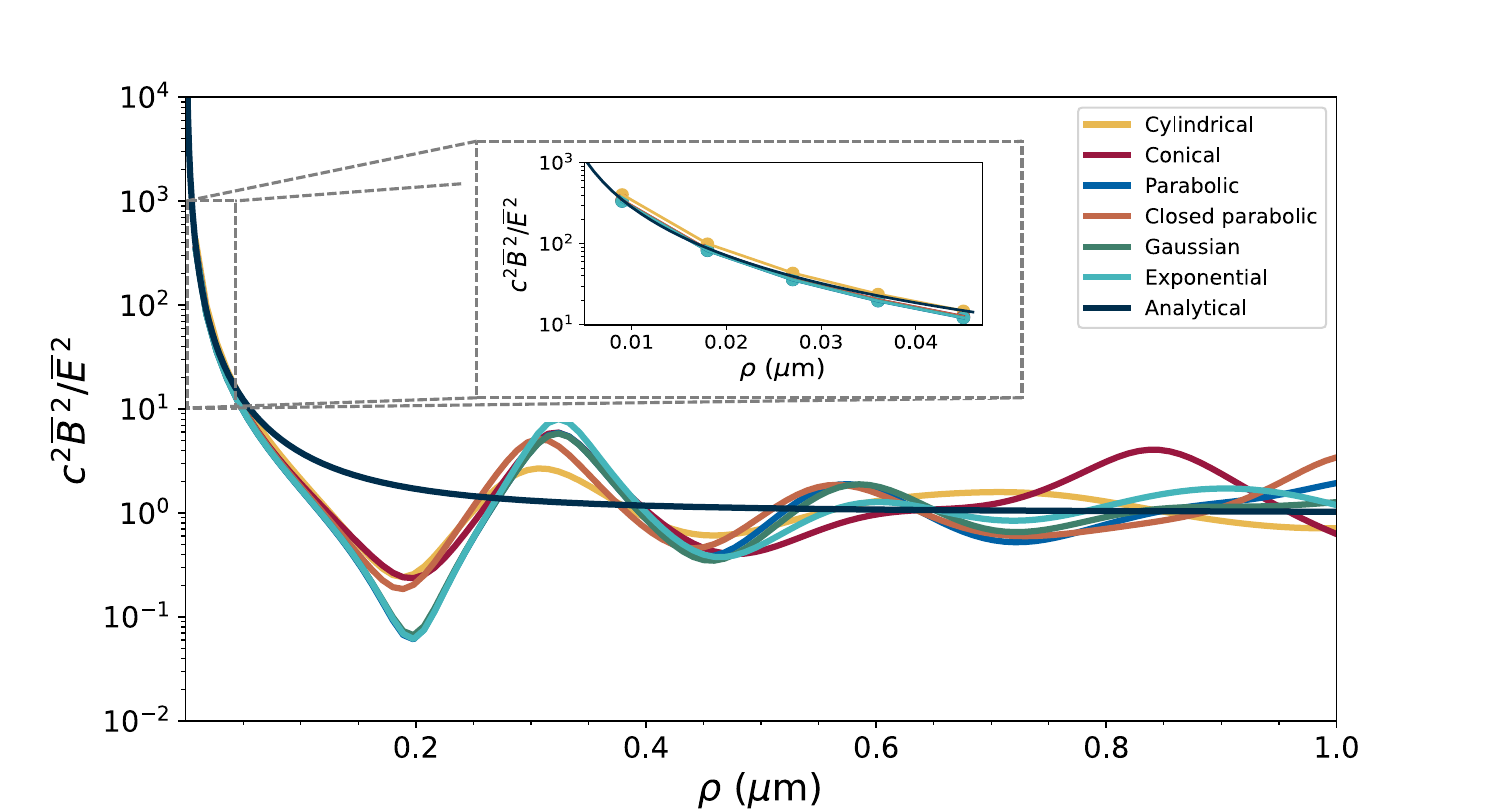}
    \caption{Transverse distribution of the intensity contrast $\frac{c^2\bar{B}^2}{\bar{E}^2}$ for different antenna geometries (see Fig.~\ref{fig:contrast_heatmap}) at the transversal plane in the center of the simulation box. 
    The abscissa shows only positive values for $y$ (equivalent to $\rho$), as the simulation obeys cylindrical symmetry in the transversal direction.}
    \label{fig:contrast_lineplot}
\end{figure}

Second and more relevant, all antennas exhibit diverging contrast near the propagation axis, as expected from the symmetry of the incident APB.
The radial extension of this central region in which the contrast is larger than 10 is $\rho_c \approx 60\,\mathrm{nm}$, whereas it remains
larger than 1 up to $\rho_c'\approx100\,\mathrm{nm}$.
The asymptotic behaviour of the intensity contrast for small $\rho$ can be derived from Eqs.~\eqref{eq:apb_efield}, for the incident beam, and \eqref{eq:apb_bfield} and assuming $\rho\ll w_0$:
\begin{linenomath}\begin{equation}
    \label{eq:analytic_contrast}
    \frac{c^2 \bar{B}^2}{\bar{E}^2} \sim 1+\left(\frac{\lambda}{\pi\rho}\right)^2.
\end{equation}\end{linenomath}
As expected, the intensity contrast exhibits a singularity at $\rho=0$, and diverges as $\rho^{-2}$. 
All simulated antennas, being cylindrically symmetric, exhibit such singularity in the intensity contrast. 
In order to see how the intensity contrast decreases along the radial coordinate for different antennas, we plot in Fig.~\ref{fig:contrast_lineplot} the intensity contrast near the beam axis for the optimized cases.
Near the propagation axis ($\rho\approx 0$), the intensity contrast follows the analytical estimation for a free APB, given by Eq.~\eqref{eq:analytic_contrast} (black line).
Away from the propagation axis, the antenna geometry imprints subtle differences.
Looking at Eq.~\eqref{eq:analytic_contrast}, the most promising way to increase the volume of the high-contrast regions is by considering longer laser wavelengths, a potential scenario for the emerging mid- and far-infrared laser systems with applications not limited to spectroscopic studies but interesting for driving and controlling magnetic samples \cite{2023_Sanchez-Tejerina}, or for photoinduced force microscopy \cite{Zeng2018}.

\section{Conclusion}

Our study provides a wide perspective for the generation of intense and isolated B-fields from structured laser beams.
We have demonstrated that the use of ultrafast APBs in tailored metallic nanoantennas provides intense longitudinal B-fields near the beam axis, highly isolated from the corresponding E-field. 
Our PIC numerical simulations show how the B-field carried by an APB can be highly amplified if using tailored antennas. 
For example, the use of conical antennas provides a gain factor up to $\sim50$ over an extended spatial region, whereas closed parabolic antennas provide a highly confined B-field with a gain factor of $\sim65$. 
This setup allows obtaining isolated B-fields of up to $\sim50$\,T from moderately intense femtosecond APBs (peak intensity of $1.9\times10^{11}\,\mathrm{W/cm^2}$) at a central wavelength of 527.5\,nm. 
It also opens a very promising route to perform MD-only optical spectroscopic studies with state-of-the-art laser technology.

We have developed an analytical model that explains the mechanism behind the B-field enhancement in terms of phase-matching between induced current loops along the propagation axis. 
The model also allows characterizing the spatial extent of the B-field needle, which expands over hundreds of nm for the parameters considered.
We have presented a systematic study using linearly sloped and curved sloped antennas, including cylindrical, conical, parabolic, Gaussian, and logarithmic shapes, showing their suitability to provide intense and highly isolated B-fields. 

Finally, we have analyzed the intensity contrast, i.e., the magnetic-to-electric field ratio, that allow us to define the isolated B-field region. 
The radial extension where the B-field intensity is larger than the E-field intensity extends up to 100\,nm, with similar results for all the explored antenna geometries. 
Note that the achieved outcomes are scalable beyond the specified central wavelength used in this work, $\lambda = 527.5\:\text{nm}$, which was picked due to its relevance in MD-only absorption spectroscopy experiments in Eu$^{3+}$ ions. 
Far-infrared and terahertz sources would allow the realization of larger isolated longitudinal B-field volumes, most suitable for a broad spectrum of technological applications ranging from  magnetic spectroscopy to controlling ultrafast magnetization dynamics.

\section*{Funding}

R.\ M.-H., L.\ S.-T., L.\ P., E.\ C.\ J., and C.\ H.-G.\ acknowledge funding from the European Research Council (ERC) under the European Union’s Horizon 2020 research and innovation program (Grant Agreement No.~851201) and support from Ministerio de Ciencia e Innovación (PID2022-142340NB-I00).
L.\ G.\ and S.\ M.\ acknowledge funding from the Austrian Science Fund (FWF) within the Young Independent Researcher Group iStOMPS (FWF~ZK-91).

\section*{Acknowledgements}

L.\ G.\ and S.\ M.\ also thank Valentina Shumakova, Elizaveta Gangrskaia, Alessandra Bellissimo, Audrius Pugzlys, and Andrius Baltu\v{s}ka for fruitful discussions and computational support by the Vienna Scientific Cluster (VSC~4). The authors thankfully acknowledge the computer resources at MareNostrum and SCAYLE, and the technical support provided by Barcelona Supercomputing Center (RES-FI-2022-3-0041).

\section*{Disclosures}
The authors declare no conflicts of interest.

\section*{Data availability}
Data underlying the results presented in this paper are not publicly available at this time but may be obtained from the authors upon reasonable request.

\clearpage
\appendix
\section{Definition of the antenna shapes}\label{apx:AntennaShapes}
We explore the longitudinal B-field amplification from azimuthally polarized beams interacting with five different antenna geometries: cylindrical, conical, parabolic, Gaussian and logarithmic. The mathematical expressions and the scanned parameters for each profile are formally given in the Table \ref{tab:geometries}. A visual sketch of the parameters in every antenna can be seen in Fig.~\ref{fig:aperture_shapes}. 
The metallic antennas are centered in the simulation box. 
Parameters $x_\text{min}$ and $x_\text{max}$ refer to the simulation box edges in the propagation axis.

\begin{table}[htb]
    \caption{Definitions of the simulated antenna shapes by analytical expressions and considered scan parameters, assuming $x \in [x_0-L/2, x_0+L/2+\Delta L]$.
    }
    \label{tab:geometries}
    \centering
    \begin{tabular}{lrlll}
        \hline
        Antenna shape &
        \multicolumn{3}{l}{Formula} &
        Scan parameters \\
        \hline 
        Cylindrical &\raggedright $\rho(x)$ &\centering=&\raggedleft$r_0$ & $L$  \\ 
        \hline
        Conical & $c_\textrm{c}$ &=&$\frac{x_\textrm{max}-x_\textrm{min}-L}{2}$   &  \\
         & $\rho(x)$ &=& $r_\textrm{max} - \frac{r_\textrm{max} - r_\textrm{min}}{L}\left(x-c_\textrm{c}\right)$ &  $r_\textrm{min}$, $r_\textrm{max}$  \\ 
         \hline
         Parabola & $f$ &=& $\frac{r_\textrm{max}^2-r_\textrm{min}^2}{4L}$ &    \\
         & $c_\textrm{P}$ &=& $\frac{r_\textrm{max}^2}{4f}$ &  \\ 
         & $\rho(x)$ &=& $2\sqrt{f\cdot(c_\textrm{P}-x)}$ & $r_\textrm{min}$, $L$, $\Delta L$  \\ 
         \hline
         Gaussian &  $\sigma$ &=& $\frac{r_\textrm{max}}{2}$  &    \\
         & $c_\textrm{G}$ &=& $\frac{x_\textrm{max}-x_\textrm{min}-L}{2}$ &   \\ 
         & $A$ &=& $\frac{r_\textrm{max}^2}{4f}$  &   \\
         & $\rho(x)$ &=& $\sqrt{-2\sigma^2\cdot\ln{\left(\frac{x-c_\textrm{G}}{A}\right)}}$ &$r_\textrm{min}$, $L$, $\Delta L$ \\ 
         \hline
         Logarithmic & $c_\textrm{e}$ &=&$\frac{x_\textrm{max}-x_\textrm{min}-L}{2}$   &  \\
          & $\rho(x) $ &=& $ r_\mathrm{max}\text{e}^{-\kappa \left(x-c_\textrm{e}\right)}$ &  $\kappa$, $L$ \\ 
        \hline
    \end{tabular}
\end{table}

\section{Analytical model}\label{apx:AnalyticalModel}

The APB will induce a set of fast oscillating current loops in the metallic antenna, and the total B-field will result from the superposition of the contributions from all these current loops. 
We start by determining the B-field created by a thin current loop, and then integrating over all the loops contained in the antenna thickness.

The induced current distribution in the aperture can be expressed as:
\begin{linenomath}
\begin{gather}\label{eq:current_dist}
    \vec{j}(\rho',\phi',x',t) = j_0(x')\delta(\rho'-\rho_0)\delta(x'-x_0)\text{e}^{\text{i}\frac{2\pi}{\lambda}(x' -  ct)}\vec{u}_\phi,
\end{gather}
\end{linenomath}
where $j_0$ is the current density amplitude, $\rho_0$ is the aperture radius and $x_0$ is the central longitudinal position of the aperture.
Strictly, this plane wave solution is not applicable for the curved wavefronts of tightly focused laser beams; still, they may be utilized in this context as first approximation.
The retarded vector potential associated to a current can be obtained, in cylindrical coordinates, as:
\begin{linenomath}
\begin{align}\label{eq:vect_pot_def}
    \vec{A}\left(\vec{r}\right) = \frac{1}{c}\int\int\int d\rho' d\phi' dx' \rho' \frac{\vec{j}(\rho',\phi',x',t_\mathrm{ret})}{|\Vec{r}-\Vec{r'}|}\vec{u}_{\phi},
\end{align}
\end{linenomath}
where we used the Coulomb gauge. 
The primed coordinates refer to the source points and $t_\mathrm{ret}=\frac{1}{c}\left(x' - |\vec{r} - \vec{r}'|\right)$, with $\vec{r}=(\rho,\phi,x)^T$, is the retarded time. 
In this way, we account for the temporal difference between the instant at which the laser beam, approximated by a plane wave, hits the respective aperture source point $r'$ and the time that the generated (electro-)magnetic fields at $r'$ take to arrive at the observer at position $r$. 
Using Eq.~\eqref{eq:current_dist} and the definition of retarded time in the previous expression we get:
\begin{linenomath}
\begin{gather}
    A_\phi\left(\vec{r}\right) = \frac{j_0(x_0)}{c}\rho_0\text{e}^{\text{i}\frac{2\pi}{\lambda}( x_0- ct)}\int d\phi' \cos(\phi')\frac{\text{e}^{\text{i}\frac{2\pi}{\lambda}\left[\rho^2+\rho_0^2 - 2\rho\rho_0\cos\phi' + (x-x_0)^2\right]^{1/2}}}{\left[\rho^2+\rho_0^2 -2\rho\rho_0\cos(\phi') + (x-x_0)^2\right]^{1/2}}.
\end{gather}
\end{linenomath}

The corresponding B-field is obtained as $\vec{B} = \nabla \times \vec{A}$. The longitudinal component of the B-field, close to the optical axis, can be approximated as:
\begin{linenomath}
\begin{align}
    B_{x,0}\left(x\right) = \lim_{\rho\rightarrow 0} (\nabla \times \vec{A})_{x} = 2\left(\frac{\partial A_\phi\left(\vec{r}\right)}{\partial \rho}\right)_{\rho=0},
\end{align}
\end{linenomath}
resulting in the following expression for the approximate longitudinal B-field created by a thin loop in the optical axis:
\begin{linenomath}
\begin{align}
    B_{x,0}\left(x\right) &= 2\pi\frac{j_0(x_0)}{c}\frac{\rho_0^2}{\rho_0^2 + (x-x_0)^2 }\left[\left(\frac{2\pi}{\lambda}\right)^2 + \frac{1}{\rho_0^2 + (x-x_0)^2}\right]^{1/2} \nonumber \\[1ex]
    &\hspace{5ex}\times \text{e}^{ \text{i} \frac{2\pi}{\lambda}\left( x_0 + \sqrt{\rho_0^2 + (x-x_0)^2} -ct\right)  - \text{i}\arctan\left(\frac{2\pi}{\lambda}\sqrt{\rho_0^2+(x-x_0)^2}\right)}.
\label{eq:one_loop}
\end{align}
\end{linenomath}
To obtain the whole longitudinal B-field created by a thick antenna, we integrate Eq.~\eqref{eq:one_loop}, which is valid for a single current loop, over the antenna thickness.
Thus, it is necessary to include the dependency of the aperture radius $\rho_0 \Rightarrow \rho_0(x')$, as well as the current density $j_0(x_0) \Rightarrow j_0(x')$  on the source position.
\begin{linenomath}
\begin{align}
    B_x\left(x\right) &= 2\pi\int dx'\frac{j_0(x')}{c}\frac{\rho_0(x')^2}{\rho_0(x')^2 + (x-x')^2 }\left[\left(\frac{2\pi}{\lambda}\right)^2 + \frac{1}{\rho_0(x')^2 + (x-x')^2}\right]^{1/2}\nonumber \\[1ex]
    &\hspace{5ex}
    \times \text{e}^{ \text{i} \frac{2\pi}{\lambda}\left( x' + \sqrt{\rho(x')^2 + (x-x')^2} -ct\right)  - \text{i}\arctan\left(\frac{2\pi}{\lambda}\sqrt{\rho(x')^2+(x-x')^2}\right)}.
    \label{eq:bx_aperture}
\end{align}
\end{linenomath}
The first factor inside the integral denotes for the current density distribution $j_0(x')$.
It is related to the electrons density and velocity by $j_0(x') = q n v(x')$, where $q$ is the electron's charge, $n$ is the density and $v(x')$ is the velocity. The electron's mean velocity can be calculated classically equaling the laser ponderomotive energy $U_p = q^2 E_0^2/4m_e\omega^2$ to the electron's kinetic energy.
\begin{linenomath}
\begin{align}
    v(x') = \frac{q E_0(x')}{\sqrt{2} m_e \omega} \Gamma(x'),  
\end{align}
\end{linenomath}
where the longitudinal change of aperture radius along the $x'$ coordinate is implemented in the E-field amplitude $E_0$. 
The $\Gamma(x')$ factor takes into account the effect of the antenna curvature and is defined as:
\begin{linenomath}
\begin{align}
    \Gamma(x') &= 1-\cos\left[\arctan\left(\frac{d}{dx'}\rho(x')\right)\right] .
\end{align}
\end{linenomath}
The second and third factors in Eq.~\ref{eq:bx_aperture} describe the decrease in B-field if the observer point $x$ is shifted away from the current loop position $x'$.
The phase factor is discussed in section \ref{sec:analytical_model}.


\bibliography{references}

\end{document}